\documentclass[conference]{IEEEtran}
\usepackage{amsmath,amssymb}
\usepackage{booktabs}
\usepackage{graphicx}
\usepackage{url}

\begin{document}

\title{voxmap-studio: an open-source speaker diarization\\
annotation tool with built-in cost instrumentation}

\author{\IEEEauthorblockN{Fumiaki Yamaguchi}
\IEEEauthorblockA{Independent Researcher\\
panchorange2203@gmail.com}}

\maketitle

\begin{abstract}
Labeling speaker diarization data is costly, yet annotation tools rarely measure
that cost. We present \emph{voxmap-studio}, an open-source, React-based
diarization annotation tool integrated with the pyannote-based diarization
ecosystem. Its canvas
is initialized by a fast stride-accelerated diarization engine so that the
annotator corrects a hypothesis rather than drawing every speaker turn by hand,
and the tool records annotation cost---typed edit-operation counts and time---as
a first-class output, enabling quantitative comparison of how much different
forms of assistance actually help. Export is gated on per-segment human
confirmation and guarded by injected ``phantom'' attention checks, which prevent
unverified automatic output from being released as ground truth. In a preliminary
study on nine AMI audio files, unassisted manual annotation was the costliest and
least accurate, and automatic initialization shifted the work from creating
turns to correcting them; highlighting uncertain segments gave the lowest cost
in our small sample. The tool and its instrumentation are open source.
\end{abstract}


\section{Introduction}
Speaker diarization---deciding ``who spoke when''---is usually solved with deep
models trained on data labeled as accurately as possible, and producing that
data is expensive, mostly in human time. Annotation tools therefore matter:
the faster and more reliably a human can turn a recording into a correct
labeling, the cheaper the data. The recent \emph{gryannote} tool~\cite{gryannote}
makes this workflow convenient by wrapping the pyannote
ecosystem in a web interface for loading a pipeline, correcting its output, and
exporting RTTM. It motivates its design by the cost of annotation, but it does
not measure that cost; how much effort each correction actually takes is left
unobserved.

We present \emph{voxmap-studio}\footnote{\url{https://github.com/panchorange/voxmap}},
an open-source diarization annotation tool that
treats annotation cost as a first-class, recorded output. The canvas is
initialized by a fast stride-accelerated diarization engine~\cite{voxmap-engine}
so that the annotator corrects a hypothesis instead of drawing from scratch, and
the interface is a hand-written React application tuned for low-level editing.
As the annotator works, the tool counts every edit operation and the time spent,
and writes them alongside the annotation, which lets us compare assistance
settings quantitatively rather than by intuition. Export is gated on per-segment
human confirmation and guarded by injected attention checks, so that automatic
output cannot leak out unverified. Our contributions are:
\begin{itemize}
\item an annotation tool that instruments annotation cost (typed edit-operation
counts and time) as a first-class output, which existing diarization labeling
tools do not record;
\item a workflow that pairs fast automatic initialization with a
confirmation-gated, attention-checked export, so that recorded annotations are
verified rather than rubber-stamped;
\item a preliminary AMI study using this instrumentation, showing that
unassisted manual annotation is the costliest and least accurate, and that
assistance shifts effort from creating turns to correcting them.
\end{itemize}

\section{The voxmap-studio tool}
\label{sec:system}

voxmap-studio is a browser-based application for producing and correcting
speaker diarization annotations. Its interface is a hand-written React
application rather than one generated from Python UI components, which lets us
tune low-level editing interactions (waveform zooming with the mouse wheel,
batch relabeling, keyboard shortcuts, and user-selectable color themes). The tool is organized around four functions:
automatic initialization, manual editing, cost instrumentation, and a
confirmation-gated export.

\begin{figure}[t]
\centering
\includegraphics[width=0.95\columnwidth]{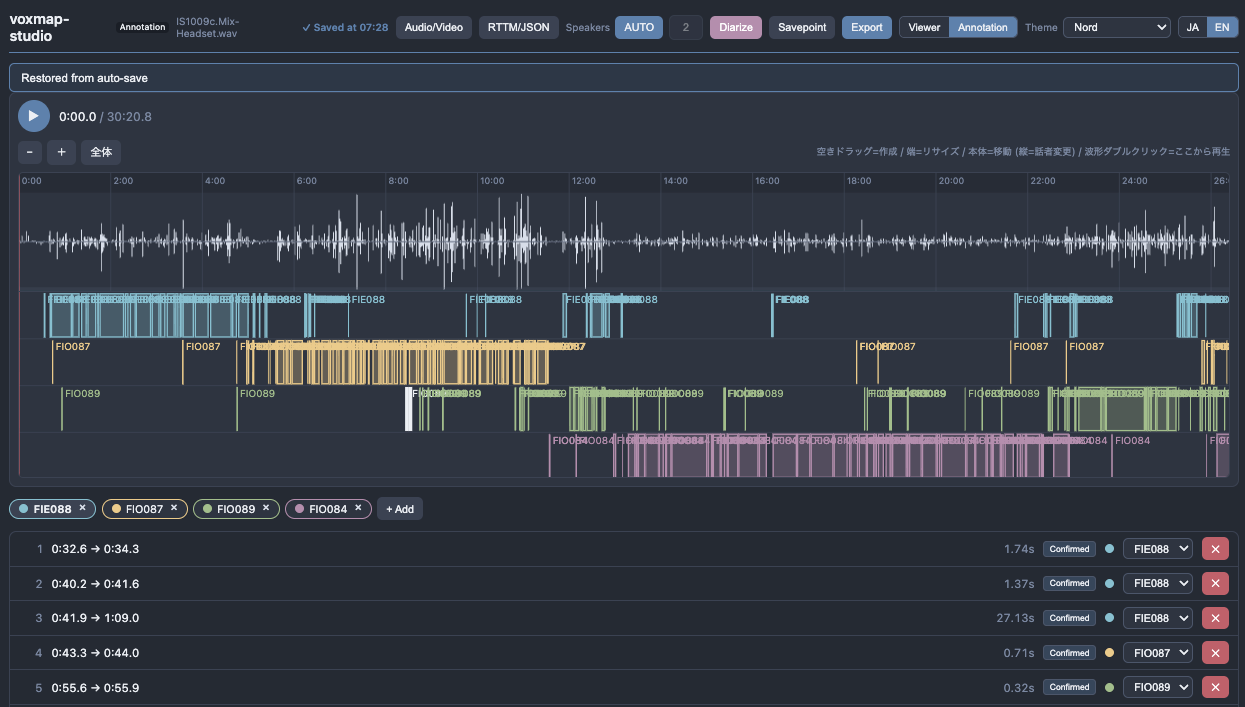}
\caption{The voxmap-studio interface: the automatic pipeline output is shown on
the waveform, edited with keyboard shortcuts and label assistance, with running
cost displayed; annotations are exported as RTTM.}
\label{fig:ui}
\end{figure}

\subsection{Automatic initialization}
To reduce the time spent creating annotations from scratch, the user can
initialize the canvas with the output of an automatic diarization
engine. We use our own stride-accelerated pipeline~\cite{voxmap-engine} built on
the pyannote~\cite{pyannote} segmentation model,
which runs at a real-time factor well below~1 on a consumer laptop, so the
first annotation appears with little waiting. The annotator then corrects this
hypothesis rather than drawing every speaker turn by hand.

\subsection{Editing and label assistance}
The annotation canvas displays speaker turns over the waveform. Each turn can be
resized, split, deleted, created, or reassigned to a different speaker, and every
operation has a keyboard shortcut. Common interactions are deliberately
lightweight: playback toggles with the spacebar, and a segment assigned to the
wrong speaker (a confusion) is reassigned to the correct one with a single number
key (\texttt{1}--\texttt{9} select a speaker and assign it to the current
selection), so the most frequent corrections need no menus.

Two optional aids speed up labeling, and both are computed from the speaker
embeddings and cluster centroids that the automatic engine already produces.
First, each segment's embedding is compared by similarity to its own speaker
centroid and to the nearest other speaker centroid: a segment that is in fact
more similar to a different speaker is flagged in red as a likely intrusion, and
a borderline segment---more similar to its own speaker, but only by a small
amount---in amber, so attention is directed to the turns most likely to be
mislabeled rather than to the whole timeline. Second, a \emph{cluster gallery}
groups candidate segments by their speaker cluster so that many turns belonging
to one speaker can be confirmed or relabeled in a single action; and pressing the
\texttt{R} key opens a candidate panel that ranks existing speakers by their
similarity to the current segment. The segment's representative embedding is
formed by averaging the pre-computed embedding vectors whose time windows overlap
the segment, so resizing the segment before pressing \texttt{R} naturally shifts
the embedding toward whatever speech the new boundaries enclose. Speaker
centroids are rebuilt on the fly from the labels the annotator has assigned so
far, and a segment is flagged as a possible new speaker when no existing centroid
is similar enough. The annotator can then reassign the segment with a click. These aids are independent and can
be turned on or off, which is what we vary in the study of Section~\ref{sec:study}.

\subsection{Built-in cost instrumentation}
A distinguishing feature of voxmap-studio is that it records the cost of
annotation as a first-class output. Every edit is counted by type
(\textsf{create}, \textsf{delete}, \textsf{split}, \textsf{resize},
\textsf{reassign}); we report their sum as \emph{edit operations}
(\textsf{editOps}), where a batch relabel of many segments counts as a single
operation, so the count reflects user gestures rather than affected segments.
The tool also records active editing time and the fraction of audio actually
listened to at normal speed. These quantities are written into a JSON sidecar
alongside the annotation, making it possible to analyze \emph{where} annotation
effort is spent and
to compare annotation conditions quantitatively---an analysis that, to our
knowledge, existing diarization annotation tools do not support.

\subsection{Confirmation-gated export}
To prevent unverified automatic output from being exported as if it were
ground truth, every segment carries a \textsf{human\_confirmed} flag. The Export
button is always available as an in-progress save, but the tool refuses to emit
the final RTTM and JSON outputs until all segments are confirmed; confirming a
segment requires having listened to its span.

To further discourage rubber-stamping, the tool can inject a small number of
\emph{phantom} segments---short fake speech turns placed in silent gaps of the
automatic output (about one per five minutes of audio, capped at eight). An attentive
annotator who listens to such a segment will find no speech and remove it; a
phantom that survives as unverified automatic output signals that the
annotator passed it through without checking. Each phantom is scored as
\emph{caught} (deleted), \emph{kept} (listened to and judged real), or
\emph{missed} (left untouched), and the counts are recorded in the sidecar.
A phantom that is left unresolved keeps the annotation in an unconfirmed state,
so the tool notifies the annotator and refuses to export until every phantom has
been dealt with; an unverified phantom therefore cannot leak into the output.

Annotations are exported in RTTM format for downstream reuse, accompanied by a
JSON sidecar that carries the per-segment \textsf{human\_confirmed} flags and the
cost counters (\textsf{editOps}, time). Each exported file additionally embeds an
integrity hash over its segments, so that any later hand-editing of the file is
detectable at evaluation time.

\section{Preliminary cost study}
\label{sec:study}

Because the tool records annotation cost directly, we can ask which forms of
assistance actually reduce it. We report a small preliminary study; it is
illustrative rather than conclusive, and we are explicit about its limits below.

\subsection{Protocol}
We annotated nine files from three AMI meetings~\cite{ami} (ES2004, IS1009,
TS3003, Mix-Headset), using the \texttt{a} segment of each meeting as a gallery
reference (available for listening and for the recommendation centroid) and
annotating the \texttt{b}, \texttt{c}, and \texttt{d} segments under three
conditions that differ in how much assistance is enabled:
\begin{itemize}
\item \textbf{C1 (manual):} no automatic initialization---the annotator draws
every speaker turn from scratch.
\item \textbf{C2 (engine + uncertainty):} the canvas is initialized with the
automatic engine and uncertain segments are highlighted, but the cluster
gallery and recommendation are not used.
\item \textbf{C3 (engine + gallery + recommend):} as C2, plus gallery-based
labeling and the recommendation function.
\end{itemize}
The three conditions and three meetings are arranged in a Latin-square fashion so
that each meeting is annotated once under each condition, giving three files per
condition. A single annotator performed all sessions. Annotations were compared
against the AMI reference with \texttt{pyannote.metrics}~\cite{pyannote-metrics}
(collar~$0.25$\,s, overlap regions skipped). Our primary cost metric is \textsf{editOps}; we also report
annotation time per audio minute as a secondary metric.

\subsection{Results}
\begin{figure}[t]
\centering
\includegraphics[width=0.92\columnwidth]{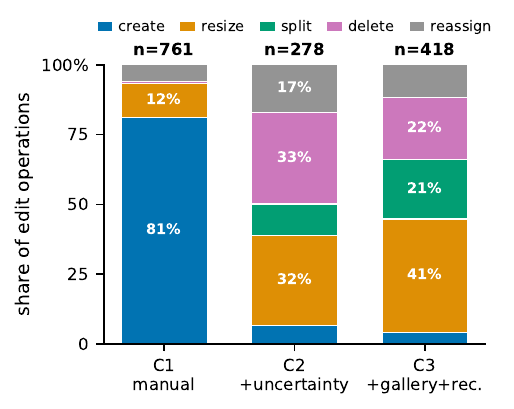}
\caption{Composition of edit operations by type for each condition, as a share
of that condition's total (mean operations per file shown as $n$ above each bar).
C1 is dominated by \textsf{create} (81\,\%), i.e.\ drawing turns from scratch;
once the canvas is initialized (C2, C3) the effort shifts to correction
operations (resize/split/delete/reassign).}
\label{fig:ops}
\end{figure}

Table~\ref{tab:main} summarizes the conditions, and Fig.~\ref{fig:ops} breaks
\textsf{editOps} down by operation type. Three observations stand out.

\emph{Unassisted manual annotation is the worst on both axes.} C1 incurs the
most edit operations (761 vs.\ 278 for C2) and also the worst error rate
(macro DER~0.177 vs.\ 0.079); its cost is dominated by \textsf{create} operations
($617$ of $761$, Fig.~\ref{fig:ops}), i.e.\ drawing turns from nothing.

\emph{Assistance changes the nature of the work.} Once the canvas is
initialized from the engine (C2, C3), \textsf{create} almost disappears and the
remaining effort shifts to correcting the hypothesis---resizing, splitting,
deleting, and reassigning.

\emph{More assistance is not monotonically cheaper.} C3, which adds gallery
labeling and recommendation, costs more operations than C2 ($418$ vs.\ $278$),
driven by a larger number of \textsf{resize} operations; the secondary time
metric follows the same ordering with a smaller spread. In this small sample,
the uncertainty-highlight condition (C2) was the cheapest and most accurate.

\begin{table}[t]
\caption{Annotation cost and quality by condition (AMI, three files each).
\textsf{editOps} is the primary metric; sec/aud-min (active editing seconds per
audio minute, idle gaps excluded) is secondary. DER$_\mathrm{mac}$ and
DER$_\mathrm{mic}$ are macro- (mean over files) and micro- (pooled over files)
averaged DER; \emph{miss} and \emph{conf.} are the missed-speech and confusion
components of the macro DER (false alarm omitted).}
\label{tab:main}
\centering
\setlength{\tabcolsep}{2pt}
\begin{tabular}{lcccccc}
\toprule
Condition & \textsf{editOps} & sec/aud-min & DER$_\mathrm{mac}$ & DER$_\mathrm{mic}$ & miss & conf.\\
\midrule
C1 manual          & 761 & 115 & 0.177 & 0.176 & 0.123 & 0.030\\
C2 +uncertainty    & \textbf{278} & \textbf{101} & \textbf{0.079} & \textbf{0.078} & 0.037 & 0.008\\
C3 +gallery+rec.   & 418 & 105 & 0.093 & 0.094 & 0.062 & 0.009\\
\bottomrule
\end{tabular}
\end{table}

\subsection{Limitations}
This study involves a single annotator and only three files per condition, so we
draw no statistical conclusions and report it as an existence proof that the
instrumentation captures meaningful differences between conditions. The annotator
had speaker-diarization annotation experience but no prior experience annotating
AMI recordings, and worked in a single loose pass---annotating each file from start
to finish without systematic re-listening or revision---so the absolute quality
is not that of a careful reference annotation. With one
file per (meeting, condition) cell, meeting difficulty and condition are not
fully separable. The reported DER measures consistency with the AMI reference,
not absolute annotation quality, and we use it only for relative comparison
across conditions. We treat time as a secondary signal and avoid conclusions
that rest on it alone.

\section{Future work}
\begin{itemize}
\item \textbf{Portable backend.} The automatic engine runs on PyTorch and is fast
when a GPU is available (CUDA on Linux/Windows, MPS on Mac), but falls back to
CPU on machines without one, where initialization is noticeably slower. Broader
adoption would benefit from faster backends on the hardware these machines
actually ship with---optimized CPU inference as well as AMD GPUs and the NPUs now
common in Windows laptops---which we plan to explore as demand warrants.

\item \textbf{Cross-session speaker assets.} C3 (gallery + recommendation) did
not outperform C2 (uncertainty highlighting alone), suggesting that the current
single-file gallery centroid does not yet exploit past annotation assets
effectively. We plan to leverage accumulated assets---gallery recordings,
previously annotated files, and corrective edits---to build richer cross-session
speaker representations, and to measure with the cost instrumentation whether
this reduces cost and improves quality beyond single-file initialization.

\item \textbf{Beyond diarization.} We aim to extend the workflow to tasks such
as speaker-attributed speech recognition.
\end{itemize}


\end{document}